\documentclass[twocolumn,showpacs,preprintnumbers,amsmath,amssymb]{revtex4}

\usepackage{graphicx}
\usepackage{dcolumn}
\usepackage{bm}
\usepackage{color}

\begin{document}


\title{Correlation and isospin dynamics of participant-spectator matter in neutron-rich colliding nuclei at 50 MeV/nucleon }

\author{Sanjeev Kumar}
\author{Y. G. Ma\footnote{Author to whom all correspondence should be addressed:
ygma@sinap.ac.cn}}
\author{G. Q. Zhang} 
\affiliation{Shanghai Institute of Applied
Physics, Chinese Academy of Sciences, Shanghai 201800, China}

\date{\today}

\begin{abstract}
The sensitivities of isospin asymmetry and collision geometry dependences of participant (overlapping region)-spectator 
(quasi-projectile and quasi-target region) matter towards the symmetry energy
using the Isospin Quantum Molecular Dynamical model are explored. 
 Particularly, the difference of number of nucleons in overlapping zone to quasi projectile-target matter is found to be 
quite sensitive towards the symmetry energy at semi-peripheral geometries compared to the individual yield. It gives us a clue of 
this quantity to act as a measure for isospin migration.  Further, the yield of neutrons (charge of the second largest fragment) 
are provided as a tool for overlapping region (quasi projectile-target) matter to check the sensitivity of above 
mentioned observable towards the symmetry energy experimentally. 
\end{abstract}

\maketitle

\section{Introduction}
The primary  goal of isospin physics is to obtain a better understanding of 
properties of asymmetric nuclear matter or in other words isospin dependence of Nuclear 
Equation of State (NEOS). Over  three decades, the 
NEOS of symmetric nuclear matter is well understood by the study of giant dipole resonances, collective flow as well as 
multifragmentation etc. \cite{Youn99,
Nato02,Dani02,Reis12,Ma}. The compressibility 
$\kappa_0~=~9\rho_0^2\frac{\partial(E/A)}{\partial\rho}|_{\rho~=~\rho_0}$, which  describes so called stiffness of the  
symmetric nuclear matter has been determined to 235 $\pm$ 14 MeV. The NEOS of  isospin asymmetric nuclear matter is recently underway, 
particularly, the density dependence of symmetry energy,
which is critically
important also in many astrophysical processes, such as, structural and dynamical 
evolution of neutron star, the critical density for the direct cooling process etc.\cite{Latti07}.

Considerable progress has been made in determining the sub- and supra-saturation density behavior of the symmetry energy 
\cite{Tsan04,Zhan12,Fami06,Tsan09,Kuma11a,Xiao09,
Russ11,Cozm11,Huang,
Gaut11}. The later part is still an 
unanswered question in spite of recent findings in term of neutron-proton 
elliptic  flow ratio and difference \cite{Russ11,Cozm11}. However, The  former one is understood to 
some extent \cite{Tsan04,Zhan12,Fami06,Tsan09,Kuma11a}, although, more efforts are needed for precise measurements.
 According to Symmetry 
Energy project (SEP) started from Michigan State University (MSU) at National Superconducting 
Cyclotron laboratory (NSCL) in 2011 (includes European as well as Asian countries), 
the sub-saturation density dependence of symmetry energy still needs more study to
verify the previously defined observables such as double ratio, isospin diffusion and then extend the study to define new 
observables to parametrize it (symmetry energy) more accurately, which has still long range in the literature from soft 
to linear one \cite{Tsan04,Fami06,Tsan09}.
Here we will try to search for new observable whose importance in 
intermediate energy heavy ion collisions is discussed below.

Recently, the role of isospin degree of freedom has been investigated using collective flow and 
its balance energy(at which flow disappears) \cite {Gaut11}. The collective flow is proven as an indicator for the 
symmetry energy. At balance energy, the attractive interactions due to mean field are balanced by the 
repulsive interactions due to the nucleon-nucleon (NN) collisions. This counterbalancing is reflected in quantities 
like participant-spectator matter \cite{Sood04}. 
Recently, the participant-spectator matter at balance energy is found to be quite 
insensitive towards the mass  and $N/Z$ of the colliding system and hence can act as a barometer for the study of 
vanishing flow \cite{Sood04,Saks12}. The inter-dependence of collective flow with symmetry energy and collective flow 
with participant-spectator matter  gives us a clue to check the sensitivity of participant-spectator 
matter towards the symmetry energy in asymmetric colliding nuclei.

Secondly, the elliptic flow is also shaped by the interplay of collisions and mean field \cite{Kuma10}. In addition, the elliptic flow 
pattern of the participant matter is affected by the presence of cold spectator matter \cite{Dani98}. 
 Specially, 
the spectator can inhibit the collective transverse expansion of the decompressing participant matter and effectively 
shadow particle emission directed towards the reaction plane. This study is  indicating that elliptical flow is influenced strongly by participant-spectator matter 
distributions. As we have discussed earlier, the elliptic flow ratio and difference from neutrons and protons was 
used as an indicator for the symmetry energy \cite{Russ11,Cozm11}. It is the another insight/clue to check the sensitivity of 
participant-spectator matter towards the symmetry energy.

Apart from collective transverse and elliptic flow, the participant-spectator matter also plays an important role in understanding 
multifragmentation as well as nuclear stopping. In recent years,
the correlation between nuclear stopping and light 
charged particles (LCPs) is investigated by using Quantum Molecular Dynamics (QMD) and Isospin-QMD (IQMD)  \cite{Dhaw06}.
 The relation is established between nuclear stopping, directed flow and 
elliptic flow \cite{Zhan07}.
Further, the participant matter is declared as an indicator to nuclear stopping \cite{Sood04}. 
All the correlation studies  indicate some indirect correlation of  participant-spectator matter with different kind of fragments. So, it also 
becomes our prime duty to correlate the participant-spectator matter with different kind of fragments and then provide 
specific type of fragments as an experimental measure for  participant-spectator matter.

 In the literature, mostly participant-spectator terminology is used, which is most suitable at higher 
incident energies, where the dynamics is almost fully accounted by NN collisions. In the energy region of present manuscript, 
the reaction proceeds through the interplay of both NN collisions and mean field. We prefer to use quasi-projectile (QP) or 
quasi-target (QT) for spectator matter and overlap region (OR) for the participant matter.

In the present work, IQMD model is used, which is discussed in detail in our recent publications \cite{Kuma11a,Kuma11},  
originally developed by  Hartnack  and collaborators \cite{Hart98}. 
The model is modified by the 
authors for the density dependence of symmetry energy, having form: 
$$E_{Sym}(\rho)~=~\frac{C_{s,k}}{2}\left(\frac{\rho}{\rho_0}\right)^{2/3}~+~\frac{C_{s,p}}{2}\left(\frac{\rho}{\rho_0}\right)^{\gamma_i},$$
with the parameters of ${C_{s,k}}$ = 25 MeV and ${C_{s,p}}$ = 35.2 MeV. When we set $\gamma_i~=~0.5$ and 1.5, respectively,  it corresponds to the soft and stiff symmetry energy
 \cite{Kuma11a}.
\section{Results and Discussion}
The present OR and QP+QT matter demonstration is based on the fireball model \cite{Gait00} as reported in Ref.\cite{Sood04}. 
All nucleons having
experienced at least one collision are supposed to originate from OR matter (labeled as $N_{OR}$). The remaining matter is called 
QP,QT matter (labeled as $N_{QP+QT}$). The $N_{Tot}$ are the total number of nucleons in the reaction system. 
This concept gives similar results as has been demonstrated in the fireball model of 
Gosset et al. \cite{Gait00} and further verified by QMD in last couple of years to till date \cite{Sood04,Saks12}.
There is another way to define the OR, QP and QT matter  on the basis of different rapidity cuts \cite{Sood04,Zhan12}.  
However, both these definitions have been reported to give the same results
\cite{Sood04,Zhan12}. In the present study, we shall use the first definition to construct the OR and QP+QT matter. 
During the expansion stage, this definition 
will lead to the production of two matters with different densities. The OR must have relatively low density compared to 
QP,QT matter. This density gradient must increase, when one goes towards the peripheral collisions. Due to the 
density gradient, the transfer of particles from high to low density region is found to relate with the phenomenon known as isospin migration \cite{Rizz08}.

Mathematically, isospin migration can be understood as:
$D^{\rho}_{n} - D^{\rho}_{p}  ~\propto ~ 4 \delta \frac{\partial E_{sym}}{\partial \rho}$,
with $D^{\rho}_{p/n}$ the  mass coefficients, which are directly 
given
by the variation of $n,p$ chemical potentials with respect to density and
asymmetry. For more detail see Ref. \cite{Rizz08}. 
From the above equation, it is clear that isospin migration 
depends on
the slope of the symmetry energy, or the symmetry pressure. When one moves towards the  semi-peripheral collisions, QP or QT of about
normal density are in contact with the OR where the density is quite low 
than saturation density.
In this region of density, a stiff symmetry energy has a smaller value but a larger 
slope in comparison with  a soft symmetry energy. In brief, this definition can help us to find the observable for isospin migration.

\subsection{Theoretical probe for isospin migration}
In the present study, thousand of events are simulated using static soft equation of state  and energy dependent NN cross-section 
for the isotopes of Sn, namely $^{112}$Sn + $^{112}$Sn, $^{124}$Sn +
$^{124}$Sn and $^{132}$Sn + $^{132}$Sn at incident energy 50 
MeV/nucleon along the whole collision geometry.

\begin{figure}
\vspace{-1.7cm}
\includegraphics[width=80mm]{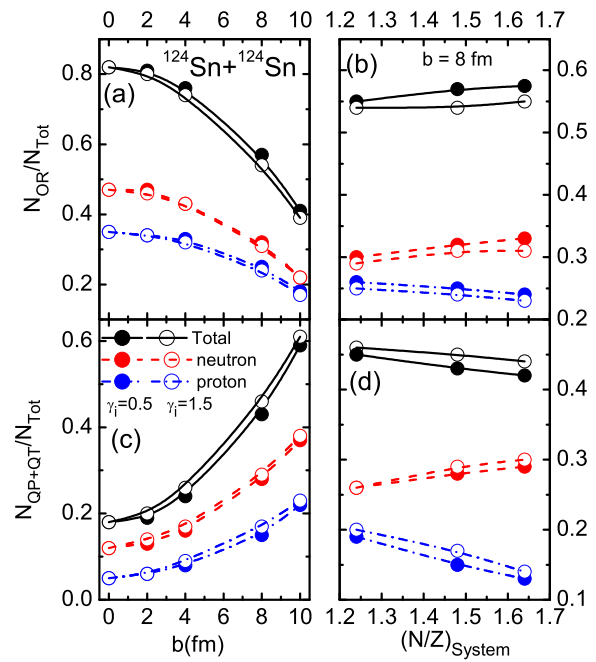}
\vspace{-1.0cm}
\caption{\label{fig:1}(Color online) Impact parameter (left panels) and isospin asymmetry (right panels) dependences of 
nucleons in OR (a,b) and QP+QT (c,d) matter including the 
contributions from neutrons and protons. The solid (open) symbols represent soft (stiff) symmetry energies. 
The OR, QP and QT are Overlapping Region, Quasi-Projectile and Quasi-Target, respectively.}
\end{figure}
 
Fig. \ref{fig:1} displays the impact parameter (b) and isospin asymmetry of the system ($(N/Z)_{system}$) dependences of 
OR and QP+QT including the neutrons and protons contribution. With the increase of impact parameter, the number of nucleons in OR (QP+QT)
matter decreases (increases). 
The similar behavior is followed by the 
impact parameter dependence of neutrons and protons. 
In contrast, 
there exists more sensitivity for the number difference between the neutrons and protons  near the central (peripheral) geometries from OR (QP+QT) matter, which 
 is due to the dominance of more NN collisions (very less collisions) near the respective collision geometries.

In semi-peripheral collisions, the isospin asymmetry dependence of OR and QP+QT matter along with neutrons and protons 
contribution have dramatic behavior. With the increase in isospin asymmetry, the number of nucleons in OR (QP+QT) matter increases (decreases).
Furthermore, the neutrons (protons) contribution from OR as well as QP+QT matter  increases (decreases).  
Interestingly, the sensitivity is stronger for QP+QT matter at higher isospin asymmetry.
This is due to the fact that 
(1) With isospin asymmetry, 
there emerges a sharp increase in neutron content inside QP+QT about 4-5 times compared to the decrease of QP+QT matter (see the red circles vs black circles in right bottom panel); 
(2) As the proton content is constant for all the isotopes of Sn and while the QP+QT matter  is decreasing, it will eventually lead to the decrease in protons content in the QP+QT matter.      
 
Lastly, the effect of symmetry energy is relatively weak on the impact parameter and isospin asymmetry dependences on
the nucleons originated from OR and QP+QT matter. This difference is true as the error bars are less than 
the size of the symbols. The soft (stiff) symmetry energy
contributes more for OR (QP+QT) matter. This is due to the gradient in  densities of two matters, which can be 
explained by slope of symmetry energy rather than magnitude.
The slope (magnitude) of symmetry energy 
changes its behavior below (at) the saturation density for the soft and stiff symmetry energy. Due to the large slope (less magnitude) for the stiff
symmetry energy even below the saturation density, 
the QP+QT is more neutron-rich with the stiff symmetry energy, which is true with the 
soft symmetry energy for OR matter as the density at freeze-out time in comparison to QP+QT is very low.   

From the above discussions, it is clear that number of nucleons contribution from OR and QP+QT matter is a good candidate to explore  the 
isospin physics, but not a potential candidate for symmetry energy (due to weak dependence). However, the density gradient of symmetry energy  gives us 
some preliminary clues of isospin migration between the nucleons of OR and QP+QT matter.

\begin{figure}
\vspace{-0.6cm}
\includegraphics[width=70mm]{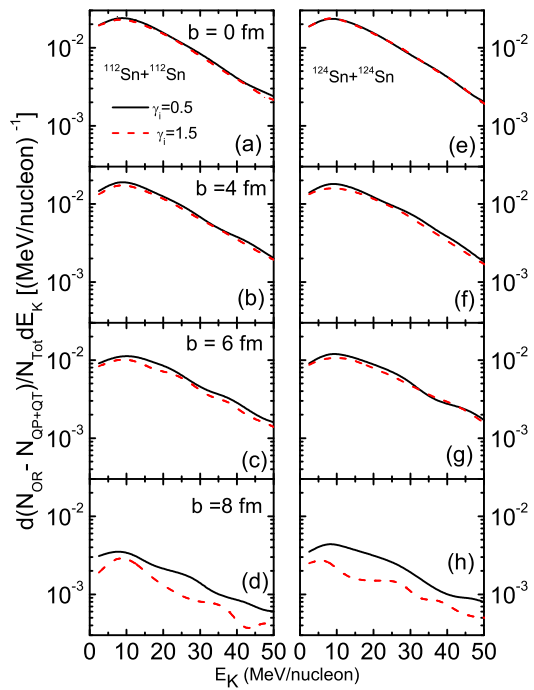}
\vspace{-0.4cm}
\caption{\label{fig:2}(Color online) The kinetic energy spectra of  the $\frac{N_{OR}-N_{QP+QT}}{N_{Tot}}$ at different colliding geometries. 
The solid (dashed) lines represent the contribution with soft and stiff symmetry energy. The a, b, c, d and e, f, g, h are for 
$^{112}$Sn+$^{112}$Sn and $^{124}$Sn+$^{124}$Sn  systems, respectively.}
\end{figure}

In order to reveal the effect of symmetry energy in term of isospin migration,  in Fig. \ref{fig:2}, we display the difference of nucleon number between OR and QP+QT ($\frac{N_{OR}-N_{QP+QT}}{N_{Tot}}$) as a function of kinetic energy of nucleons at 
different impact parameter and for two neutron-rich 
systems. In central collisions, 
no effect of symmetry energy 
is observed. However, with the increase of impact parameter, sensitivity of symmetry energy towards the 
observable seems increasing. This indicates that the effect of symmetry energy on 
the observable under study depends weakly on the isospin asymmetry, but strongly on the density gradient. As discussed earlier, 
the density gradient  increases strongly (weakly) with the impact parameter (isospin asymmetry). 
The effect of symmetry energy on the observable ($\frac{N_{OR}-N_{QP+QT}}{N_{Tot}}$) mainly 
originates from the density difference and not due to the isospin asymmetry. 
We further  know, the density difference is a direct measure of the isospin migration.
The findings are  also  supporting the results of the Ref.~\cite{Rizz08}. 

From the study, we  can conclude that the difference of number of nucleons from OR to QP+QT  can act as a 
probe to the slope of symmetry energy versus the density in term of isospin migration at sub-saturation 
densities. 
\subsection{Correlation with fragments and observables for experiments}

The second aspect of this paper is to correlate the number of nucleons in OR and QP+QT matter with fragmentation process 
and then provide some observables to test the above observable as a
sensitive probe for symmetry energy. 

\begin{figure}
\vspace{-2.0cm}
\includegraphics[width=75mm]{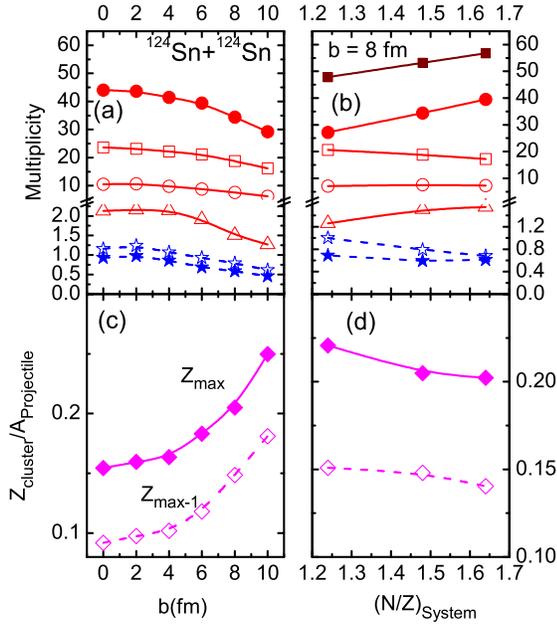}
\vspace{-1.0cm}
\caption{\label{fig:3}(Color online) Same as Fig. \ref{fig:1}, but for the multiplicity of different kind of fragments from lighter to the heavier ones. 
 In upper panels (a,b), the solid squares represent for free nucleons, solid circles (open squares) for free neutrons (protons), 
circles (triangles) for deuteron (triton), open stars (solid stars) for $^3$He ($^4$He). In lower panels (c, d), 
solid and open diamonds for the heaviest fragment and second heaviest fragment, respectively.
}
\end{figure}

To this end, we displayed the impact parameter and isospin asymmetry dependences 
of different kind of fragments in Fig. \ref{fig:3}, which is just like in Fig. \ref{fig:1}.  The multiplicity of different kind of light charged particles goes on decreasing with the increase of 
impact parameter as well as  size of the fragments. This type of the behavior has been already observed many times in 
the literatures \cite{Dhaw06}.  On the other hand, $Z_{max-1}/A_{projectile}$
(charge of the second largest fragment normalized by the mass number  of projectile) and $Z_{max}/A_{projectile}$ (charge of the  largest fragment normalized by the mass number  of projectile)  increase with 
impact parameter. These two L. H.S. panels  are behaving exactly the similar
just like the L. H.S. panels in Fig. \ref{fig:1}, indicating some clues of 
the correlation for the measure of number of nucleons in  OR(QP+QT) using the light charged particles (charge of the heavier fragments).

\begin{figure}
\vspace{-2.0cm}
\includegraphics[width=80mm]{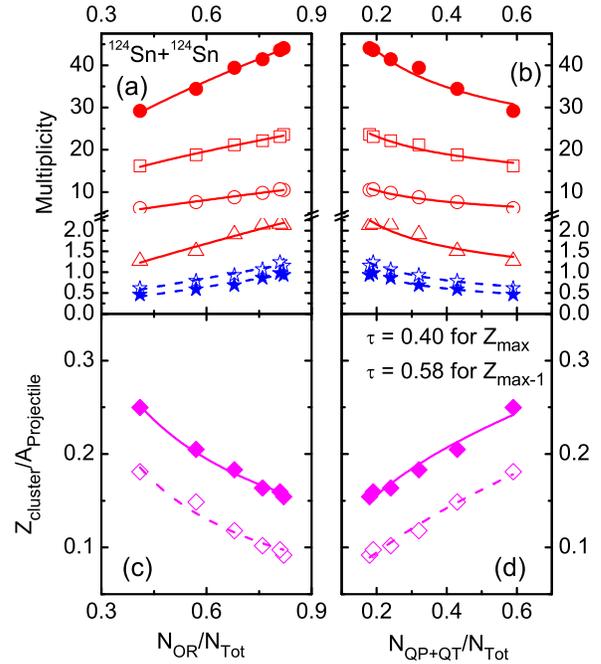}
\vspace{-1.0cm}
\caption{\label{fig:4}(Color online) Correlation between the nucleons of OR and QT+QT matter with different kind of fragments. 
The lines are fitted with the power-law  form $Y=CX^{\tau}$. Symbols are the same as Fig. 3.}
\end{figure}

To further strengthen the correlation, the R.H.S. of Fig. \ref{fig:3} give us some interesting 
features. Just like the isospin asymmetry
dependence of nucleons in OR (QP+QT) matter, the multiplicity of free nucleons (charge of the heavier fragments) is increasing 
(decreasing). 
Moreover, with the isospin asymmetry, the increasing (decreasing) trend of neutrons (protons) contribution in 
OR matter (Fig.\ref{fig:1}) is similar to the behavior of the multiplicity of neutrons (protons) (Fig.\ref{fig:3}). 
In addition, the isospin 
asymmetry dependence of charges of heavier fragments is also similar to the  contribution of 
protons from the QP+QT (Fig. \ref{fig:1}). These are the 
strong evidences for correlation. The important point to check here is the sensitivity of 
yield of  different kind of fragments and charge of the heavier fragments
with nucleons of OR and QP+QT matter, which will be further checked in Fig. \ref{fig:4} by using the 
power-law fitting method. 

From the light charged particles, interesting isospin effects are observed with isospin asymmetry of the system. 
With the increase in the one neutron in the LCPs, i.e. from $^1$H$\rightarrow$$^2$H$\rightarrow$$^3$H and $^3$He$\rightarrow$$^4$He, the multiplicity
changes its trend from decreasing $^1$H,~$^3$He to increasing for $^3$H,~$^4$He, respectively. 
However, with the increase of one proton from $^3$H to $^3$He, the 
multiplicity reveals a sharp change from increasing to decreasing. The neutrons-protons distributions within the LCPs is also
satisfying the neutrons-protons distributions from the OR zone. This  indicates the strong correlation between LCPs and 
OR matter, but the point of interest is which type of LCPs are perfect indicator for OR matter  which will be discussed in Fig.\ref{fig:4}.

\begin{table}[t]
\caption{$\tau$ value extracted from Fig.\ref{fig:4}, showing the positive (negative) slope of light charged particles with 
participant (spectator) matter.}\begin{center}
\begin{tabular}{c  c  c  c  c  c  c}
\hline
\hline
Particles           &~~~$n$ &~~~~$p$ &~~~~$^2$H &~~~~$^3$H &~~~~$^3$He &~~~~$^4$He \\
\hline 
$\tau$ for OR        & 0.605   & 0.546   &0.815      & 0.831     &  1.022     &  1.180    \\
$\tau$ for QP+QT        & -0.317   &  -0.288  &   -0.423   &    -0.418  &  -0.524     &  -0.599\\
\hline
\hline
\end{tabular}
\end{center}
\vspace{-0.9cm}
\label{tautable}
\end{table}

In Fig. \ref{fig:4}, the sensitivity of number of nucleons in
OR and QP+QT matter with the multiplicity of LCPs as well as charge of the heavier fragments is checked by fitting the 
power law of the form $Y=CX^{\tau}$. The positive correlation is observed between OR (QP+QT) nucleons and 
multiplicity of LCPs ($Z_{max}~ and~ Z_{max-1}$). 
Interestingly, from the power-law slope, it is found that although, the multiplicity decreases with the size of light fragments,
the slope parameter or sensitivity increases with the fragment size (shown in Table \ref{tautable}). This type of prediction is 
also true for the second largest fragment 
and  the largest fragment
for the QP+QT nucleons.   
The slope parameter for neutrons (0.6) with 
nucleons of OR matter (also shown in  Table \ref{tautable}) and for $Z_{max-1}/A_{Projectile}$ with 
nucleons of QP+QT matter 
(0.58) (Fig.\ref{fig:4}) is almost the same, which reveals the similar sensitivity of the respective matter towards the respective fragments. From here, one can say that if one uses the neutron as a measure for the nucleons of OR matter and $Z_{max-1}/A_{Projectile}$ as a 
nucleons of QP+QT matter, then the difference of two parameters at semi-peripheral geometry, just like in Fig. \ref{fig:2},
can act as a probe for the isospin migration or slope of symmetry energy. 
From the Table \ref{tautable}, 
we also observed that there is a linear correlation
between the nucleons of OR matter and $^3$He fragments (i.e., having $\tau$ very close to one), which  indicates that the $^3$He 
can be taken as a direct
measure for the nucleons of OR matter.

\section{Conclusion}
In summary,  we tried to 
find an observable to measure density dependence of symmetry energy or its slope by studying the OR and QP+QT matter in 
intermediate energy heavy-ion collisions. 
The difference in number of nucleons of OR to QP+QT matter is particularly sensitive towards the slope 
of symmetry energy at semi-peripheral geometries. This gives us a clue that this observable can act as probe for the isospin migration. 
Principally, the yield of neutrons and  charge of the second largest fragment ($Z_{max-1}/A_{Projectile}$)
could be provided as  experimental tools  to check the sensitivity of nucleons in OR and QP+QT matter towards the symmetry energy in term of isospin migration. 
Interestingly, the nucleons of OR matter has a linear correlation with the yield of $^3$He ($\tau$ close to 1). 

\section*{ACKNOWLEDGMENTS}
This work is supported
in part by the Chinese Academy of Sciences Fellowships for young
international scientists under the Grant No. 2010Y2JB02, and  by
NSFC of China under contract No.s 11035009, 10979074, the Knowledge
Innovation Project of CAS under Grant No. KJCX2-EW-N01.

\end{document}